\documentclass[aps,prb,twocolumn,showpacs,superscriptaddress,amsmath,amssymb,floatfix,longbibliography,10pt]{revtex4-2}

\usepackage{graphicx}
\usepackage{amssymb}
\usepackage[toc,page]{appendix}
\usepackage{color}
\usepackage{comment}
\usepackage{amsmath}
\usepackage{dcolumn}
\newcolumntype{d}[1]{D{.}{.}{#1}}
\usepackage{array}
\usepackage{siunitx}
\usepackage{hyperref}
\usepackage{epsf}
\usepackage{float}
\usepackage{longtable}
\usepackage{booktabs}
\usepackage{threeparttable}

\newcommand{\R}{\mathbf{r}}

\newcommand{\UP}{n_{\uparrow}}
\newcommand{\DN}{n_{\downarrow}}
\newcommand{\TUP}{\tau_{\uparrow}}
\newcommand{\TDN}{\tau_{\downarrow}}

\newcommand{\be}{\begin{equation}}
\newcommand{\ee}{\end{equation}}
\newcommand{\bea}{\begin{eqnarray}}
\newcommand{\eea}{\end{eqnarray}}
\newcommand{\bean}{\begin{eqnarray*}}
\newcommand{\eean}{\end{eqnarray*}}

\newcommand{\myr}[1]{{ #1}}

\newcommand{\Dabel}[1]{\label{#1}}

\begin{document}

\title{Restoring the Point-and-Charge Gradient Expansion for the \\
Strong Interaction Density Functionals}
\author{Lucian A. Constantin}
\affiliation{Institute for Microelectronics and Microsystems (CNR-IMM),
Via Monteroni, Campus Unisalento, 73100 Lecce, Italy}
\author{F. Naeem}
\affiliation{Center for Biomolecular Nanotechnologies, Istituto Italiano
di
Tecnologia, Via Barsanti 14, 73010 Arnesano (LE), Italy}
\affiliation{Dipartimento di Matematica e Fisica, Universit{\`a} del Salento, Via per Monteroni, 73100 Lecce, Italy}
\author{E. Fabiano}
\affiliation{Institute for Microelectronics and Microsystems (CNR-IMM), Via
Monteroni, Campus Unisalento, 73100 Lecce, Italy}
\affiliation{Center for Biomolecular Nanotechnologies, Istituto Italiano
di
Tecnologia, Via Barsanti 14, 73010 Arnesano (LE), Italy}
\author{F. Sarcinella}
\affiliation{Institute for Microelectronics and Microsystems (CNR-IMM), Via
Monteroni, Campus Unisalento, 73100 Lecce, Italy}
\affiliation{Center for Biomolecular Nanotechnologies, Istituto Italiano
di
Tecnologia, Via Barsanti 14, 73010 Arnesano (LE), Italy}
\author{Fabio Della Sala}
\affiliation{Institute for Microelectronics and Microsystems (CNR-IMM), Via
Monteroni, Campus Unisalento, 73100 Lecce, Italy}
\affiliation{Center for Biomolecular Nanotechnologies, Istituto Italiano di
Tecnologia, Via Barsanti 14, 73010 Arnesano (LE), Italy}

\date{\today}
\def\zPC{{ePC\,}}
\begin{abstract}
The strong-interaction functionals $W_\infty[n]$ and $W'_\infty[n]$ play an important role in the adiabatic-connection method of Density Functional Theory.
The strictly-correlated electron approach can be used to exactly compute these
functionals, yet calculations are computationally very expensive even for small electronic systems, and thus  semilocal approximations have been proposed.
In this work we develop a meta-generalized gradient approximation (meta-GGA) model for the strong-interaction functionals, enhanced point-and-charge (\zPC), constructed from exact constraints. 
In particular, the \zPC restores the second-order gradient expansion of the PC model, that is relevant for the equilibrium properties of Wigner crystals, and ensures the non-negativity of $W'_\infty[n]$. 
We assess the \zPC model for atoms and various model systems: Hooke's atoms, two-electron exponential densities, $s$- and $p$-hydrogenic shells, quasi-two-dimensional infinite barrier model, perturbed uniform electron gas and $H_2$ dissociation. We prove a good overall accuracy of the \zPC model, that achieves a broader applicability than any previous semilocal models. 
\end{abstract}

\pacs{71.10.Ca,71.15.Mb,71.45.Gm}

\maketitle

\section{Introduction}
\Dabel{sec1}
The Kohn-Sham (KS) Density Functional Theory (DFT) \cite{hohenberg1964inhomogeneous,kohn1965self} 
is the most used theoretical method in electronic structure calculations 
\cite{burke12,book2,burke12,giustino_book,becke14,jones15}. The KSDFT success is based on its simplicity and accuracy, 
using only the electronic density $n(\R)$ as the basic variable which 
determines all the ground-state properties of the electronic system \cite{kohn1996density,kohn1999nobel}. 

The accuracy of the KSDFT is directly dependent on the approximation used for the exchange-correlation (XC) 
energy functional \cite{scuseriaREVIEW05}. In the KSDFT adiabatic connection (AC) method 
\cite{langreth1975exchange,gunnarsson1976exchange,savin2003adiabatic,
cohen2007assessment,ernzerhof1996construction,burke1997adiabatic,colonna1999correlation}, the XC energy functional $E_{xc}[n]$ is defined as 
\begin{eqnarray}
E_{xc}[n]&=&\int_0^1 d\alpha\;W_{xc,\alpha}[n], \\
W_{xc,\alpha}[n]&=&\langle\Psi_n^{min,\alpha}|\hat{V}_{ee}|\Psi_n^{min,\alpha}\rangle-U[n],
\Dabel{eqI1}
\end{eqnarray}
where $U[n]=(1/2)\int d\R\int d\R'\;n(\R)n(\R')/|\R-\R'|$ is the Hartree energy,
$\hat{V}_{ee}$ is the Coulomb repulsion operator, and
$\Psi_n^{min,\alpha}$ is the antisymmetric wave function that yields the density $n(\R)$ and
minimizes the expectation value $\langle\hat{T}+\alpha \hat{V}_{ee}\rangle$.
\myr{Thus, $\Psi_n^{min,\alpha}$ is the ground-state wave-function of an $N$-electron system with the Hamiltonian \cite{gori2009electronic,seidl07,seidl2000density,seidl2000densitye}
\begin{eqnarray}
\hat{H}_\alpha[n]&=&\hat{T}+\alpha\hat{V}_{ee}+\hat{V}^\alpha_{ext}[n],\nonumber\\
\hat{V}^\alpha_{ext}[n]&=&\sum_{i=1}^N v_{ext}^\alpha ([n],\R_i),
\label{eqI1new1}
\end{eqnarray}
where $\hat{V}^\alpha_{ext}$ is the external potential that ensures $\hat{H}_\alpha[n]$ has the same ($v$-representable) density $n(\R)$ for any $\alpha$. 
}
Here 
$\alpha\ge 0$ is the coupling constant, also known as the interaction strength.

The AC integrand $W_{xc,\alpha}[n]$ is exactly known in the weak- ($\alpha\rightarrow 0$) and 
strong- ($\alpha\rightarrow \infty$) interaction limits \cite{seidl2000simulation,gorling1993correlation,jana2020generalizing,liu2009adiabatic}:
\begin{eqnarray}
W_{xc,\alpha\rightarrow 0}[n]&=& W_0[n]+ \sum_{m=2}^\infty m E_c^{GLm}\alpha^{m-1}, \\
W_{xc,\alpha\rightarrow\infty}[n]&=&W_\infty[n]+W'_\infty[n]\alpha^{-1/2} + \nonumber \\
&& W^{(2)}_\infty[n]\alpha^{-3/2}+\dots,
\Dabel{eqI2}
\end{eqnarray}
Here, $W_0[n]=E_x^{EXX}[n]$ is the exact DFT exchange functional and $E_c^{GLm}$ is the $m$-th term of the 
G\"{o}rling-Levy (GL) perturbation theory \cite{gorling1994exact,gorling1993correlation,gorling1995hardness}.
In the strong-interaction limit, the strictly correlated electron (SCE) approach 
\cite{gori2009density,malet2012strong,friesecke2022strong} becomes exact, and the square of the SCE wave function for a $N$-electron 
system is \cite{gori2009electronic}
\begin{eqnarray}
&& |\Psi_{SCE}(\R_1,\dots,\R_N)|^2=\frac{1}{N!}\sum_P \int d\R \frac{n(\R)}{N}\delta(\R_1-f_{P(1)}(\R)) \nonumber \\
&& \times \delta(\R_2-f_{P(2)}(\R))\dots \delta(\R_N-f_{P(N)}(\R)),
\Dabel{eqI3}
\end{eqnarray}
where $f_1, \dots,f_N$ are comotion functions (with $f_1(\R)=\R$), and
$P$ is a permutation of ${1,\dots, N}$. 
\myr{ In the SCE limit, the electrons are well separated from each other, such that the effect on the total energy of the spin state is usually neglected \cite{seidl2000density,seidl2000densitye,gori2009electronic,seidl07}, being of order $\mathcal{O}(e^{-\alpha^{1/4}})$ \cite{gori2009electronic}.}

\myr{ We recall that the set of comotion functions $\{f_i(\R)\}$ represents the electron configurations that yield the global (and degenerate) minimum of the asymptotic potential-energy function 
\begin{eqnarray}
E_{pot}([n],\{\R_i\})&=&\lim_{\alpha\rightarrow\infty}\frac{\hat{H}_\alpha[n]}{\alpha}=
\hat{V}_{ee}+\sum_{i=1}^{N}v_{SCE}([n],\R_i),\nonumber\\
v_{SCE}([n],\R_i)&=&\lim_{\alpha\rightarrow\infty}\frac{v_{ext}^\alpha([n],\R_i)}{\alpha},
\label{eqInew3}
\end{eqnarray}
such that in the SCE limit, the potential energy function $E_{pot}([n],\{f_i(\R)\})$ becomes 
\begin{equation}
E_{SCE}=W_\infty[n]+U[n]+\sum_{i=1}^{N}v_{SCE}([n],f_i(\R)).
\label{eqInew4}
\end{equation}
For large, but finite $\alpha\gg 1$, the electrons execute small zero-point oscillations around the SCE configuration $\{f_i(\mathbf{s})\}$, such that \cite{gori2009electronic,seidl07}
\begin{eqnarray}
E_{pot}([n],\{\R_i\})&=&E_{SCE}+\frac{1}{2}\sum_{\mu,\nu=1}^{3N}\frac{\partial^2 E_{pot}
}{\partial r_\mu \partial r_\nu}\biggm|_{\{\R_i\}=\{f_i(\mathbf{s})\}} \nonumber\\
&& \times
(r_\mu-f_\mu(\mathbf{s}))(r_\nu-f_\nu(\mathbf{s}))+....
\label{eqInew5}
\end{eqnarray}
}

Thus, the strong-interaction functionals in the SCE approach are \cite{gori2009electronic}
\begin{eqnarray}
&& W_\infty[n]=\int d\R \sum_{i=1}^{N-1}\sum_{j=i+1}^N \frac{n(\R)}{N|f_i(\R)-f_j(\R)|}-U[n] , \\
&& W'_\infty[n]=\frac{1}{4N}\int d\R n(\R) \sum_{\mu=4}^{3N}\omega_\mu(\R) > 0, 
\Dabel{eqI4}
\end{eqnarray}
where $\omega_\mu(\R)$ are the frequencies of the local normal modes around the degenerate SCE minimum (i.e. the eigenvalues of 
the Hessian of the total energy, \myr{ see Eq. (\ref{eqInew5})}).

The Adiabatic Connection Integrand Interpolation (ACII) XC functionals represent approximations of
$W_{xc,\alpha}[n]$, usually constructed from interpolating between the weak- and strong-interaction limits 
\cite{seidl1999strictly,seidl2000simulation,seidl2000density,seidl2000densitye,
perdew2001exploring,magyar2003accurate,
gori2009electronic,liu2009adiabatic,liu2009adiabatic2,
sun2009extension,
seidl2010adiabatic,gori2010density,mirtschink2012energy,zhou2015construction,
vuckovic2016exchange,fabiano2016interaction,vuckovic2017interpolated,vuckovic2017simple,
giarrusso2018assessment,
vuckovic2018restoring,kooi2018local,constantin2019correlation,fabiano2019investigation,
smiga2020modified,smiga2022selfconsistent,
vuckovic2023density,constantin2023adiabatic,jana2023semilocal,genisi2}.
Thus, the ACII XC functionals are non-linear functions of several functionals:
\begin{equation}
E_{xc}^{ACII}[n]=
\mathcal{F}(W_0[n],E_c^{GL2}[n],W_\infty[n],W'_\infty[n]),
\Dabel{eqI5}
\end{equation}
where only the first two terms of the weak- and strong-interaction series (see Eq. (\ref{eqI2})) are considered. 

Because the SCE approach is computationally very expensive, the ACII XC functionals use semilocal approximations for 
$W_\infty[n]$ and $W'_\infty[n]$. Here we mention the point-charge-plus-continuum (PC) model \cite{seidl2000density,seidl2000densitye},
its revision revPC \cite{gori2009electronic}, the modified PC (mPC) \cite{constantin2019correlation} and 
the harmonium PC (hPC) \cite{smiga2022selfconsistent}. 

The exact PC model has the following GE2 expressions \cite{seidl2000density,seidl2000densitye}
\begin{eqnarray}
W_\infty^{PC-GE2}[n]&=&\int d\R A n^{4/3}(\R)F^{PC-GE2}(s), \\
W_\infty^{'PC-GE2}[n]&=&\int d\R C n^{3/2}(\R)F^{'PC-GE2}(s), \\
F^{PC-GE2}(s)&=&1-\mu_\infty s^2, \\
F^{'PC-GE2}(s)&=&1+\mu'_\infty s^2,
\Dabel{eqne1}
\end{eqnarray}
where $\mu_\infty=0.14$ and $\mu'_\infty=0.491$. 
Here $A=-1.451$ and $C=1.535$, while $s=|\nabla n|/[2(3\pi^2)^{1/3}n^{4/3}]$ is the reduced gradient of the density. 
Note that the exact value $\mu'_\infty=0.491$ was 
not considered in any previous PC model; in place of it,  negative values (usually $ \mu'_\infty= - 0.7222$) were used \cite{seidl2000density,seidl2000densitye,smiga2022selfconsistent,gori2009electronic}, which have been obtained from considerations of atomic/spherical systems \cite{seidl2000density,seidl2000densitye,gori2009electronic}.

The coupling-constant integrand can be also found 
from any given XC functional, using the relation \cite{seidl2000density,seidl2000densitye,gorling1993correlation}
\begin{equation}
W_{xc,\alpha}^{app}[\UP,\DN]=\frac{d}{d\alpha}\Bigg(\alpha^2 E_{xc}^{app}[n_{\uparrow,1/\alpha},n_{\downarrow,1/\alpha}] \Bigg),
\Dabel{eqI6}
\end{equation}
where $n_{\sigma,1/\alpha}=(1/\alpha^3)n_\sigma(\R/\alpha)$ is the uniformly scaled spin-density.
Taking the low-density limit $\alpha\rightarrow \infty$ in Eq. (\ref{eqI6}), one can extract 
the $W_{\infty}^{app}[\UP,\DN]$ and $W_{\infty}^{'app}[\UP,\DN]$.
Thus, commonly used approximations are the local spin density (LSD) $W_{\infty}^{LSD}[\UP,\DN]$ and $W_{\infty}^{'LSD}[\UP,\DN]$, 
the Perdew-Burke-Ernzerhof (PBE) \cite{perdewPRL96} generalized gradient approximation (GGA) 
$W_{\infty}^{GGA}[\UP,\DN]$ and $W_{\infty}^{'GGA}[\UP,\DN]$ and the meta-GGA models 
$W_{\infty}^{MGGA}[\UP,\DN]$ and $W_{\infty}^{'MGGA}[\UP,\DN]$ \cite{perdew2004meta,seidl2000density,seidl2000densitye}. 
These model approximations for $W_\infty[n]$ and $W'_\infty[n]$ have been tested for various systems 
\cite{seidl2000density,seidl2000densitye,gori2009electronic,constantin2019correlation,smiga2022selfconsistent,jana2023semilocal}.

The construction of accurate $W_\infty[n]$ and $W'_\infty[n]$ is of interest not only for the 
ACII functionals, but also for the DFT development in the low-density regime, e.g. by implying an inverse 
engineering technique from Eq. (\ref{eqI6}). In fact, the low-density limit is rich on various phenomena, 
such as the uniform electron gas (UEG) ﬁrst-order phase transition from paramagnetic to body-centered cubic
crystal at bulk parameter  $r_s = 86.6(7)$ \cite{UEGlowden1}, 
the UEG soft plasmon mode \cite{PerdewPNAS2021,PerdewCPkernelnew1}, or the classical asymptotic 
behavior of XC potential far from metal surfaces in the vacuum \cite{PitarkePRL25}.
Finally, we recall that the second-order gradient expansion (GE2) coefficient of the correlation energy
is known only in the high-density limit \cite{bruecknerPR68}, and at the metallic densities (where 
$2\le r_s\le 6$) \cite{huPRB86}, but not in the low-density regime where it was approximated from
the LSD linear response \cite{jelliumLR1}. In this sense, the gradient expansion of the PC model can be of high interest.  

In this work, we construct a meta-GGA model for the strong-interaction functionals $W_\infty[n]$ and $W'_\infty[n]$,
named enhanced  PC (\zPC), 
that restores the original GE2 of the PC model, and solves several
shortcomings of the existing models. 
In particular, previous models do not guarantee that $W_\infty[n]<0$ and $W'_\infty[n]>0$, under all density conditions.

\myr{There are not many available benchmark SCE references for $W_\infty[n]$ and $W'_\infty[n]$. In this respect, we report several SCE calculations for two-electron systems and an accurate estimation for quasi-two-dimensional infinite barrier model that may help to judge on the accuracy of the models. Nevertheless, the meta-GGA models for the strong-interaction limit can be further improved considering additional non-local ingredients, such as the Yukawa potential \cite{yukxc}.
}

 The paper is organized as follows: In Sec. II, we present and explain the
construction of the \zPC model. In Section III, we report the results for atoms and various model systems: Hooke's atoms, two-electron model densities, $s$- and $p$- hydrogenic shells,  quasi-two-dimensional infinite barrier model, perturbed uniform electron gas and $H_2$ dissociation; in Sec. IV we present molecular systems, considering for atomization energies and ionization potentials.
Finally, in Sec. V, we summarize our conclusions, and in Sec. VI we provide the computational details. 

\section{Theory}
\Dabel{sec2}


\subsection{The \zPC model for $W_\infty$:}
\Dabel{sec221}
A meta-GGA expression for $W_\infty$ can be written as:
\begin{equation}
W_\infty^{\zPC}[n]=\int d\R A n(\R)^{4/3}F^{\zPC}(s(\R),z(\R)),
\Dabel{eqne2}
\end{equation}
where \myr{ $A=-1.451$} and $z=\tau^W/\tau$ is the well-known meta-GGA ingredient \cite{perdew2004meta,meta16},
with 
 $\tau=\sum_{i=1}^N |\nabla\phi_i|^2/2$ and $\tau^W=|\nabla n|^2/(8n)$ being the positive-defined KS kinetic energy density and the von Weizs\"{a}cker kinetic energy density, respectively. (Here $\phi_i$ is the $i$-th occupied KS orbital). We recall that $z$ is always bounded between 0 and 1, 
 $z=1$  for \myr{ the ground-state of} any one- and two-electron \myr{ (closed-shell)} systems, while for a slowly
varying density regime, $z\approx (5s^2/3)/\Big(1+(5s^2/3) \Big)\rightarrow 0$ \cite{perdew2004meta,meta16}. \myr{ Note that Eq. (\ref{eqne2}) is correctly independent of the spin polarization, see the discussion after Eq. (\ref{eqI3}).}

For $F^{\zPC}(s,z)$ we propose the form
\begin{eqnarray}
F^{\zPC}(s,z)&=& F_0(s)+ \Big( zF_1(s)-F_0(s)\Big)z^p,\\
F_1(s)&=& a_1+\frac{a_2}{1+a_3 s^8}, \\
F_0(s)&=& 1-\kappa + \frac{\kappa}{1+\frac{\mu_\infty s^2}{\kappa}+\frac{\mu_\infty^2 s^4}{\kappa^2}},
\Dabel{eqne3}
\end{eqnarray}
where $F_1(s)$ is the value of $F^{\zPC}$ when $z=1$ (i.e. for one and two-electron systems) and $F_0(s)$ is the value of $F^{\zPC}$ when $z=0$ (e.g. in the UEG),  and $p$ is a parameter to be fixed. 

\myr{ Note that $F_0(s)$ and $F_1(s)$ are simple GGA expressions: $F_0(s)$ is constructed to be accurate for slowly-varying density regions, while $F_1(s)$ should describe one- and two-electron systems as well as iso-orbital regions.  
Eq. (19) represents a simple interpolation between these limits, where the parameter $p$ models the strength of the connection between these limits (i.e. when $p\to 0,\; z\ne 0$ then $F^{\zPC}\to zF_1(s)$ and when $p\to \infty,\;z\ne 1$ then $F^{\zPC}\to F_0(s)$). Quite similar expressions have been considered in meta-GGA exchange functionals using other iso-orbital indicator e.g. \cite{mgga_ms0}.   
}

The parameters of $F_1(s)$  have been fitted to best reproduce the SCE values of all one and two-electron atoms in Tab. \ref{tab1} \myr{ (i.e. H, He, two-electron exponential density and three Hooke's atoms)},
obtaining $a_1=0.1$, $a_2=0.9342$, $a_3=0.22447$.
\myr{The term $s^8$ in the denominator ensures that $F_1$ is a constant at small $s$ ($F_1\approx a_1+a_2$ for $s\le 0.7$) and approaches fast the asymptotic behavior ($F_1\approx a_1$ for $s\ge 2$). We have found these features to be important for the accuracy of the \zPC model for the two-electron systems described in Sec. \ref{sec3}
}
Note that for one-electron systems, $W_\infty[n]=E_x^{EXX}=-U$, while 
for many-electron systems 
\begin{equation}
-U[n] \le W_\infty[n] \le E_x^{EXX}[n],
\Dabel{eqne4}
\end{equation}
where the first inequality occurs from Eq. (\ref{eqI4}).

On the other hand, $F_0(s)$ 
dominates for slowly-varying densities (i.e. when $s$ and $z$ are small). 
The parameter $\kappa=1-A_x/A=0.491$ has been fixed from the condition that 
$A_x/A \le F_0(s) \le 1$, where $A_x=-\frac{3}{4\pi}(3\pi^2)^{1/3}$ is the LDA exchange prefactor. 
This condition will ensure that, in the slowly-varying density limit, 
$W_\infty^{\zPC}[n] \le E_x \approx E_x^{LDA}$. Thus, $F_0(s)$ is 
non-empirical, and when $s\rightarrow 0$, it behaves as
\begin{equation}
F_0(s)\rightarrow 1-\mu_\infty s^2+\mathcal{O}(s^6),\;\;\;\rm{when}\;\;\; s\rightarrow 0,
\Dabel{eqne5}
\end{equation}
such that the PC-GE2 is accurately recovered.

%
\begin{figure}[h]
\includegraphics[width=\columnwidth]{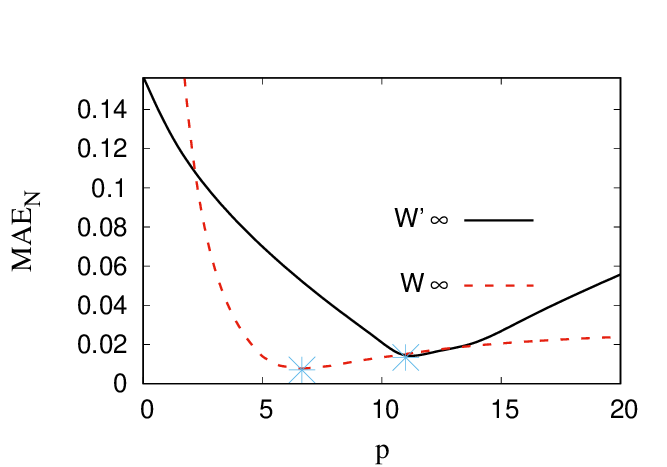}
\caption{The mean absolute error per electron $MAE_N$ for $W_\infty^{\zPC}(p)$ (red curve) and $W_\infty^{'\zPC}(p)$ (black curve),
as a function of the parameter $p$. The minimum error is at $p=6.65$ for $W_\infty^{\zPC}$ and 
at $p=11$ for $W_\infty^{'\zPC}$, respectively.
}
\Dabel{f1}
\end{figure}
%
Finally, in order to find the parameter $p$, we consider eight atoms for which the $W_\infty^{SCE}$ values are known \myr{ \cite{seidl07,daas2023lo}} (i.e. Li, Be, B, C, Ne, Ar, Kr and Xe),
and we compute the mean absolute error per electron $MAE_N (p)= \frac{1}{8}\sum_{i=1}^8 |W_\infty^{SCE}-W_\infty^{\zPC}(p)|/N_i$,
where $N_i$ is the number of electrons of the $i$-th atom. 
We plot in Fig. \ref{f1}, $MAE_N$ versus $p$, and we take $p=6.65$, where the error is minimum. This choice completes 
the functional construction for $W_\infty^{\zPC}$.

\subsection{The \zPC model for $W'_\infty$:}
\Dabel{sec222}
We consider the functional approximation for $W'_\infty$ of the form
\begin{equation}
W_\infty^{'\zPC}[\UP,\DN]=\int d\R\; C n^{3/2}(\R)F^{'\zPC}(s,z,\zeta),
\Dabel{eqne6}
\end{equation}
where \myr{ $C=1.535$}, $\zeta=(\UP-\DN)/n$ is the relative spin polarization \cite{perdew2004meta}, and 
\begin{eqnarray}
F^{'\zPC}(s,z,\zeta)&=& F'_0(s)+ \Big( z^pF'_1(s,\zeta)-F'_0(s)\Big)z^2,\nonumber\\
F'_1(s,\zeta)&=& \Big( b_1+\Big[b_1+b_2s^2 \Big]e^{-b_3s^6}\Big)\Big( 1-\zeta^{10}\Big),\nonumber\\
F'_0(s)&=& \frac{1+(\mu'_\infty +1)s^2}{1+s^2}.
\Dabel{eqne7}
\end{eqnarray}
Similarly to the case of $W_\infty$, $F'_1(s,\zeta)$ is constructed from two-electron systems,
obtaining the parameters $b_1=0.04865$, $b_2=4.3217$ and $b_3=16.581$. 
Note that for any one-electron system, $F'_1(s,\zeta=1)=0$ such that $W_\infty^{'\zPC}[\UP,\DN]$
is exact. By construction, the term $\Big( 1-\zeta^{10}\Big)$ is practically important only for $\zeta \ge 0.7$, 
otherwise $F'_1$ can be considered independent of spin polarization.

For the slowly-varying density regime, $F'_0(s)$ dominates by construction, 
and recovers the PC-GE2 as
\begin{equation}
F'_0(s)\rightarrow 1+\mu'_\infty s^2+\mathcal{O}(s^4),\;\;\;\rm{when}\;\;\; s\rightarrow 0.
\Dabel{eqne8}
\end{equation}
To find the parameter $p$, we use the same technique as before. Thus, 
we consider 5 atoms for which $W_\infty^{'SCE}$ is known \myr{ \cite{gori2009electronic}} (i.e. Li, Be, B, C, and Ne)
and we compute $MAE_N (p)= \frac{1}{5}\sum_{i=1}^5 |W_\infty^{'SCE}-W_\infty^{'\zPC}(p)|/N_i$,
showing it in Fig. \ref{f1}. The error is minimum at $p=11$, that completes
the functional construction for $W_\infty^{'\zPC}$.

\subsection{Properties of the \zPC model:}

The \zPC model satisfies the following properties:
\begin{itemize}
\item \myr{ $F^{\zPC}(s,z)$ and $F^{'\zPC}(s,z,\zeta)$ are analytical functions and do not have any order-of-limits problem (e.g. $\lim_{z\to 1}\lim_{s\to 0}F^{\zPC}=\lim_{s\to 0}\lim_{z\to 1}F^{\zPC}=F_1(0)$). We recall that meta-GGA functionals that depend on both meta-GGA ingredients $z$ and $\alpha=(\tau-\tau^W)/\tau^{unif}$, (e.g. TPSS \cite{TPSS} and TM \cite{TMmeta} exchange functionals) can have this problem that may cause severe failures \cite{orderlimit1,orderlimit2,orderlimit3}. 
Note that in this work we use the bounded iso-orbital indicator $z$ to fulfill the known constraints on $W_\infty$ and $W'_\infty$. For properties of other iso-orbital indicators, see Refs. \cite{isoindicators1,isoorbital2}. 
}

\item Under a uniform density scaling $n_\lambda(\R)=\lambda^3 n(\lambda \R)$, we have \cite{seidl2000density,seidl2000densitye}
\begin{eqnarray}
&& W_\infty^{\zPC}[n_\lambda]=\lambda W_\infty^{\zPC}[n],\nonumber\\
&& W_\infty^{'\zPC}[n_\lambda]=\lambda^{3/2} W_\infty^{'\zPC}[n];
\Dabel{eqne9}
\end{eqnarray}

\item For a slowly-varying density region, where $z\approx (5/3)s^2/[1+(5/3)s^2]$, the \zPC model recovers the PC-GE2 \myr{ (see Eqs. (13)-(16)):}
\begin{eqnarray}
&& W_\infty^{\zPC}[n]\rightarrow  \int d\R\; A n^{4/3}(\R) \big( 1 - \mu_\infty s^2 + \mathcal{O}(s^6)\big),\nonumber\\
&& W_\infty^{'\zPC}[n]\rightarrow  \int d\R\; C n^{3/2}(\R) \big( 1 + \mu'_\infty s^2 + \mathcal{O}(s^4) \big).\nonumber\\
\Dabel{eqne10}
\end{eqnarray}
\myr{ We recall that $W_\infty^{PC}$ \cite{seidl2000density,seidl2000densitye}, $W_\infty^{mPC}$ 
\cite{constantin2019correlation}, $W_\infty^{hPC}$ \cite{smiga2022selfconsistent} also behave as $W_\infty^{PC-GE2}$ at small reduced gradients $s$. On the other hand,
$W_\infty^{'PC-GE2}$ is recovered in the slowly-varying density limit only by $W_\infty^{'\zPC}$, while all the other known models ( $W_\infty^{'PC}$, $W_\infty^{'mPC}$, $W_\infty^{'hPC}$, $W_\infty^{'MGGA}$ of Ref. \cite{seidl2000density,seidl2000densitye}) fail to fulfill this exact constraint. 
}

Instead, for a constant density, all these models yield the LDA behavior: $W_\infty^{LDA}[n]=\int d\R \;A n^{4/3}(\R)$
and $W_\infty^{'LDA}[n]=\int d\R \;C n^{3/2}(\R)$;

\item   
The negativity and positivity conditions, respectively:
\begin{eqnarray}
&& W_\infty^{\zPC}[n] \le  0, \nonumber \\
&& W_\infty^{'\zPC}[n] \ge  0 .
\Dabel{eqne11}
\end{eqnarray}
Note that PC and hPC models fail for both of these conditions, while mPC and meta-GGA models satisfy them.
Moreover, the tighter bound $W_\infty[n] \le E_x^{EXX}[n]$ still holds for mPC and meta-GGA models, while 
by construction $W_\infty^{\zPC}[n] \le E_x^{LDA}[n]$.

\item
In the strong-interaction limit of a many-electron system (where in most energetic regions $z\le 0.5$), 
the electrons are usually well separated, such that they perform zero-point oscillations around equilibrium 
positions in a Wigner crystal \cite{seidl2000density,seidl2000densitye,seidl07,friesecke2022strong}. 
Consequently, in this case, both $W_\infty$ and $W'_\infty$ must be spin-independent.

On the other hand, in the strong-interaction limit of a few-electron systems (where $z\ge 0.5$), 
the spin dependence can still be important, for example, to correctly describe spin-polarized small
systems, as H and Li atoms. However, \myr{ $W_\infty[n]$ can be constructed as a functional of the total density, being spin-independent, but still accurate for H and Li atoms (as reported in Table \ref{tab1} for PC \cite{seidl2000density,seidl2000densitye}, hPC \cite{smiga2022selfconsistent} and \zPC models)}. On the other hand, $W'_\infty$ should vanish for H atom, and this condition requires a spin-dependence of the functional 
(i.e. $W'_\infty[\UP,\DN]$). Then the \zPC model satisfies  
\begin{eqnarray}
&& W_\infty^{\zPC}[\UP,\DN]=W_\infty^{\zPC}[\UP+\DN],\;\;\; {\rm for}\;\;\; 0 \le z\le 1 \nonumber\\
&& W_\infty^{'\zPC}[\UP,\DN]\approx W_\infty^{'\zPC}[\UP+\DN], \;\;\; {\rm for}\;\;\; z \le 0.5. \nonumber\\
\Dabel{eqne12}
\end{eqnarray}
These conditions are violated by the meta-GGA models \cite{seidl2000density,seidl2000densitye,perdew2004meta}.

\item \zPC is exact for the H atom (where $z=1$ and $\zeta=1$)
\begin{eqnarray}
&& W_\infty^{\zPC}=-0.3125, \nonumber\\
&& W_\infty^{'\zPC}=0.
\Dabel{eqne13}
\end{eqnarray}
We recall that for any one-electron system, 
$W_\infty=E_x^{EXX}=-U$. Even if this fully-nonlocal constraint cannot be recovered by semilocal functionals \cite{umetax}, the meta-GGA approximations can be accurate for many one-electron systems. On the other hand, $W'_\infty=0$, an exact condition that is fulfilled by the \zPC model.

\end{itemize}

%
\begin{figure}[h]
\includegraphics[width=\columnwidth]{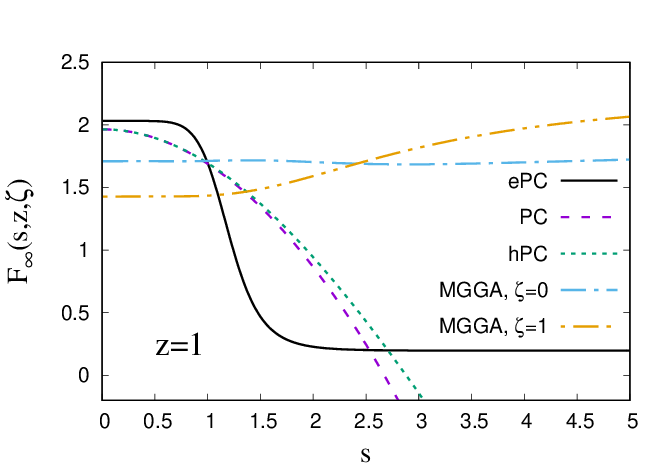}
\includegraphics[width=\columnwidth]{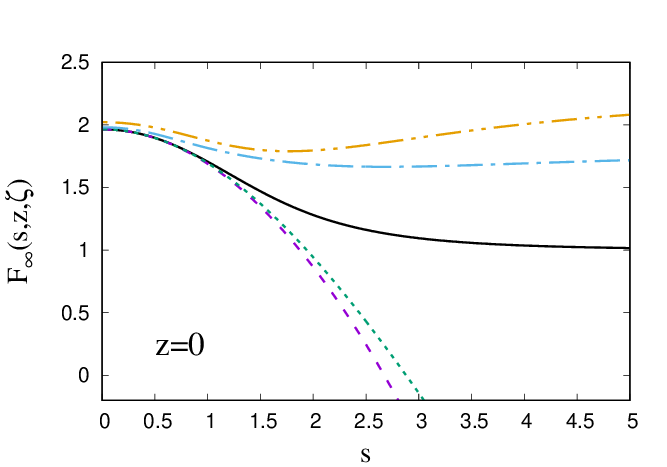}
\caption{The enhancement factor $F(s,z,\zeta)$ of the $W_\infty$, for
$z=1$ (upper panel) and	$z=0$ (lower panel) versus the reduced gradient	$s$, 
in case	of several models (\zPC, PC, hPC and meta-GGA).
}
\Dabel{f2}
\end{figure}
%
%
\begin{figure}[h]
\includegraphics[width=\columnwidth]{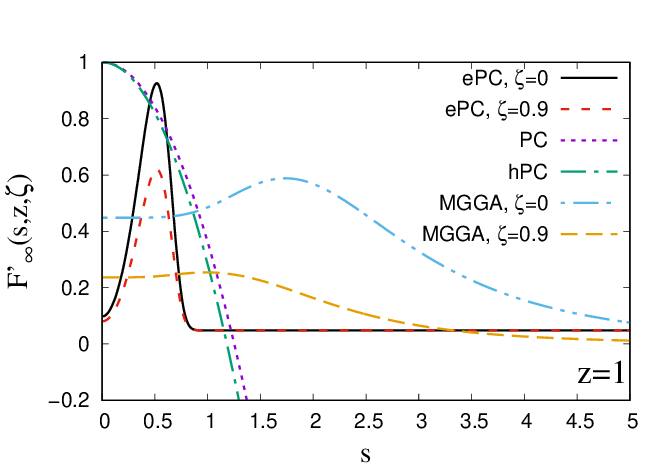}
\includegraphics[width=\columnwidth]{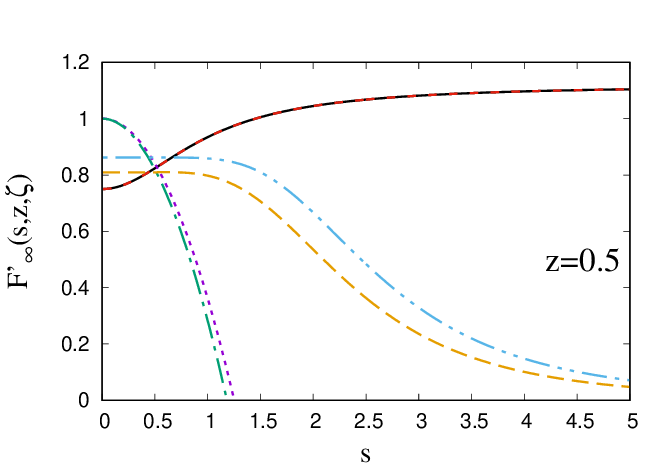}
\includegraphics[width=\columnwidth]{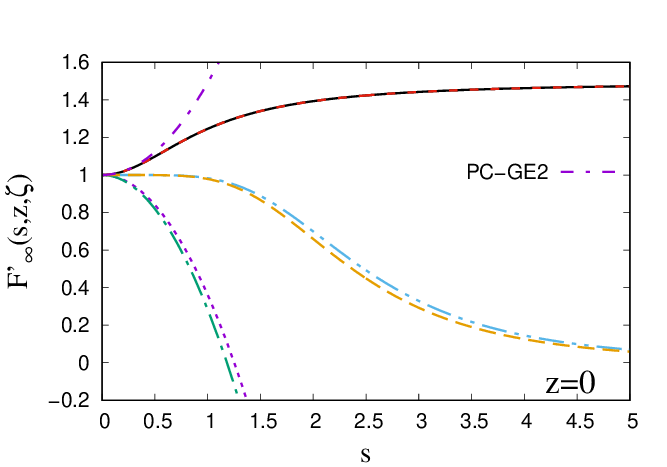}
\caption{The enhancement factor $F'_\infty(s,z,\zeta)$ of the $W'_\infty$, for
$z=1$ (upper panel), $z=0.5$ (middle panel) and $z=0$ (lower panel) versus the reduced gradient $s$, 
in case of several models (\zPC, PC, hPC and meta-GGA).
}
\Dabel{f3}
\end{figure}
%
Now, let us consider the following definitions for the enhancement factors
\begin{eqnarray}
&& F_\infty(s,z,\zeta)=\frac{w_\infty(\UP,\DN,\nabla\UP,\nabla\DN,\TUP,\TDN)}{e_x^{unif}(n)}\nonumber\\
&& F_\infty'(s,z,\zeta)=\frac{w'_\infty(\UP,\DN,\nabla\UP,\nabla\DN,\TUP,\TDN)}{w_\infty^{'unif}(n)},
\Dabel{eqne14}
\end{eqnarray}
where $e_x^{unif}(n)=A_x n^{4/3}$ is the exchange energy density of a spin-unpolarized uniform electron gas, and $w_\infty^{'unif}(n)=C n^{3/2}$.

In Fig. \ref{f2}, we show a comparison between the enhancement factors of several models for $W_\infty$.
We observe that $F_\infty^{PC}(s)$ and $F_\infty^{hPC}(s)$ are positive for $s \lesssim 2.68$ and 
$s \lesssim 2.88$, respectively, while for larger values of reduced gradients, both $F_\infty^{PC}(s)$ and $F_\infty^{hPC}(s)$ 
are negative. On the other hand, $F_\infty^{\zPC}(s,z)$ and $F_\infty^{MGGA}(s,z,\zeta)$ are positive for any $s$, $z$, and $\zeta$. 
In case of $z=1$ (upper panel), $F^{\zPC}\approx a_1$ for $s\ge 2.2$, but when $z$ decreases,
$F^{\zPC}$ becomes more uncompressed, as shown for the $z=0$ (lower panel) where $F^{\zPC}\approx 1$ for $s\ge 4$. 
Finally, we mention that $F^{MGGA}$ is significantly dependent on $\zeta$, for any value of $z$. This failure is not 
present in the other models. 

In Fig. \ref{f3}, we show the enhancement factors of several models for $W'_\infty$.
Now, $F^{'PC}(s)$ and $F^{'hPC}(s)$ are positive only for $s \lesssim 1.25$ and
$s \lesssim 1.18$, respectively. Regarding the \zPC model, we observe that 
at $z=1$ it starts from $2b_1(1-\zeta^{10})$ at $s=0$, showing a sharp peak (whose magnitude is $\zeta$-dependent) at $s\approx 0.5$, and at $s\gtrsim 0.9$ rapidly approaches the constant asymptote $b_1$.  

On the other hand, for $z=0.5$ (middle panel) and $z=0$ (lower panel), $F^{'\zPC}(s,z,\zeta)$ is almost
independent of the relative spin-polarization, by construction, correcting the behavior of the meta-GGA. 
Note that at $z=0$, $F^{'\zPC}(s,z,\zeta)$ recovers PC-GE2 for $s\lesssim 0.7$ and increases monotonically from 1 to $1+\mu'_\infty$, being very different from all the other models.

\section{Applications to atoms, model systems and H$_2$ dissociation:}

\Dabel{sec3}
\subsection{Atoms}
\Dabel{sec31}
%
\begin{table}[ht]
\caption{\Dabel{tab1} $W_\infty$ and $W'_\infty$ (in a.u.) for several atoms
and spherical model systems (Hooke's atom with three 
values of $\omega$: 0.5, 0.1, and 0.036) \cite{smiga2022selfconsistent} and the two-electron exponential density $n(r)=2/\pi e^{-2r}$ (Exp.).
For atoms we use the EXX orbitals and densities, while for the 
model systems we use the exact orbitals and densities. Last lines of each panel show the mean absolute error (MAE)
and the mean absolute error per electron MAE$_N$.
Best results are in boldface.
}
\begin{tabular}{lrrrrr}
\hline \hline
  & SCE & PC & hPC & \zPC & TPSS  \\
 \hline\hline
\multicolumn{6}{c}{$W_\infty$} \\
H & -0.3125 & -0.313 & -0.329 & {\bf -0.3125} & {\bf -0.3125} \\ 
Hook(0.5) & -0.743 & -0.702  & {\bf  -0.743} & -0.758 & -0.754  \\
Hook(0.1) & -0.304 & -0.284 & {\bf -0.303}  & -0.311 & -0.308  \\
Hook(0.036) & -0.170 & -0.156 & -0.167 & -0.174 & {\bf -0.170}  \\
Exp. & -0.910 & -0.886 & -0.906 &{\bf  -0.913} & -0.917  \\
He & -1.500 & -1.463  & -1.492  & {\bf -1.498}  & -1.512  \\
Li & -2.603 & -2.558  & -2.582  & {\bf -2.600}  & -2.590  \\
Be & -4.021  & -3.943 &-3.976   & {\bf -4.020} & -3.980  \\
B & -5.706 & -5.643  & {\bf -5.676}  & -5.756  & -5.660  \\
C & -7.782 & -7.714  & {\bf -7.753}  & -7.853  & -7.710  \\
Ne & -20.035 & -20.018 & -20.079  & {\bf -20.035} & -19.979  \\
Ar & -51.555 & {\bf -51.547} & -51.615 & -51.191 & -51.380  \\
Kr & -166.850 & -167.356 & -167.438 & -166.539 & {\bf -166.776}  \\
Xe & -322.835   & -324.521 & -324.619 & -323.346 & {\bf -323.344}  \\
\hline
MAE &-& 0.186&	0.188 &	0.096 &	{\bf 0.073} \\
MAE$_N$ &- & 0.0125 & 0.0079 & {\bf 0.0055} & 0.0057    \\
 \hline\hline
\multicolumn{6}{c}{$W'_\infty$} \\
H & {\bf 0} & 0.043 & 0.025 & {\bf 0} & {\bf 0} \\
Hook(0.5) & 0.208 & 0.215  &{\bf 0.208}  & 0.215 & 0.240  \\
Hook(0.1) & 0.054 & {\bf 0.054} & 0.053  & 0.053 & 0.062  \\
Hook(0.036) & 0.022 & {\bf 0.021} &  {\bf 0.021} & 0.020 & 0.026  \\
Exp. & 0.293 & 0.344 & {\bf 0.308} & 0.333 & 0.339  \\
He & 0.621 & 0.729  & 0.646   & {\bf 0.636}  & 0.728  \\
Li & 1.38 & 1.622  & \bf{1.434}   & 1.448 & 1.528  \\
Be & 2.59   & 2.919  & {\bf 2.600}   & 2.624 & 2.713  \\
B & 4.2 & 4.694  & \bf{4.244}   &  4.270 & 4.405  \\
C & 6.3   & 7.079  & 6.475   & {\bf 6.429} & 6.683  \\
Ne & 22  & 24.425  &  23.045  & {\bf 21.997} & 23.835  \\
Ar & - & 86.17 & 82.59  & 79.854  & 86.76  \\
Kr & - & 381.01 & 370.34 & 376.26 & 390.99  \\
Xe & -   & 886.3 & 866.05 & 894.27 & 914.2  \\
\hline
MAE  & - & 0.407 &	0.127 &	{\bf 0.034} & 	0.263 \\
MAE$_N$ & - & 0.069 & 	0.019 & 	{\bf 0.009} &	0.042\\
%
\hline \hline
\end{tabular}
\end{table}
%
%
\begin{figure}[h]
\includegraphics[width=\columnwidth]{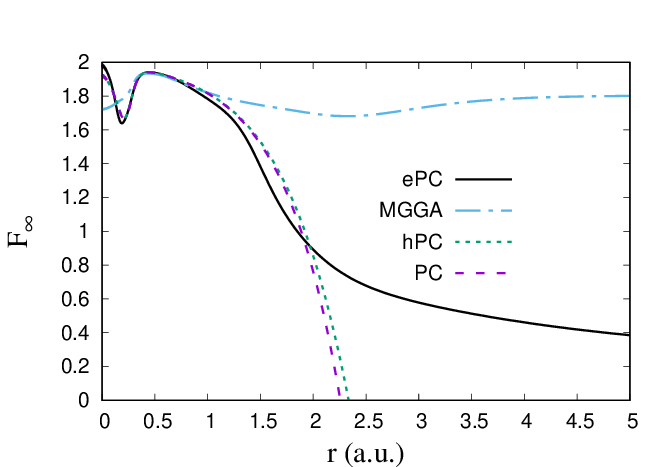}
\includegraphics[width=\columnwidth]{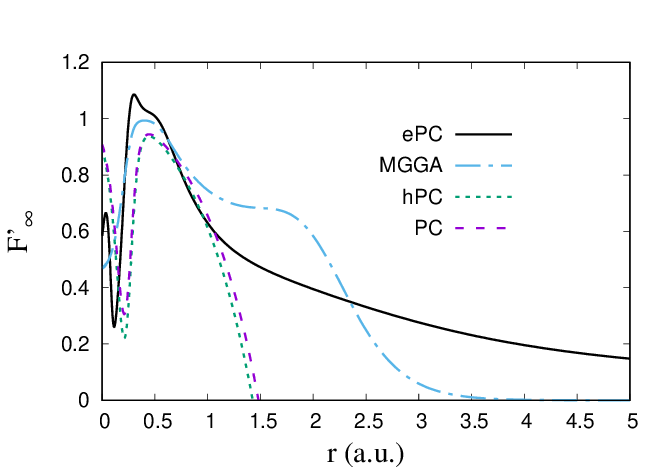}
\caption{The enhancement factors $F_\infty$ (upper panel) and $F'_\infty$ (lower panel) 
versus the radial distance from the nucleus $r$, for the Ne atom.
}
\Dabel{f4}
\end{figure}
%
In Table \ref{tab1}, we report the $W_\infty$ and $W'_\infty$ results for atoms. 
The SCE reference values were taken from Refs. \cite{smiga2022selfconsistent,gori2009electronic,seidl07,jana2023semilocal,
genisi2,daas2023lo}. 
$W_\infty^{\zPC}$  and $W_\infty^{TPSS}$ give the best results 
(with MAE$_N$=5.5 $\cdot 10^{-3}$ and 5.7 $\cdot 10^{-3}$)
closely followed by $W_\infty^{hPC}$, while $W_\infty^{PC}$ has larger errors.
Nevertheless, we mention that $W_\infty^{PC}$ is significantly better than $W_\infty^{mPC}$, $W_\infty^{PBE}$ or $W_\infty^{LDA}$, see Refs. \onlinecite{seidl2000density,seidl2000densitye,jana2023semilocal}. 
\myr{ We also note that for heavier atoms (Ar, Kr, and Xe), all functionals are less accurate, $W_\infty^{TPSS}$ yielding the best performance. In particular, for the Ar atom, $W_\infty^{\zPC}$ gives the worst result, with MAE$_N$=0.02. 
}

On the other hand, in case of $W'_\infty$, \zPC (MAE$_N$=9$\cdot 10^{-3}$) is by far  the most accurate, with a MAE$_N$ 2-7 times better than all other
methods.
\myr{For He to Ne atoms, we see that the PC and TPSS overestimate while hPC and \zPC are significantly more accurate. 
For the heavier atoms (Ar, Kr, Xe), the differences between the model predictions are quite substantial, such that SCE benchmark calculations are needed to understand the performance of these models for heavy atoms.}

In order to visualize the \zPC behavior, we also show in Fig. \ref{f4}, the enhancement factors $F_\infty$ 
and $F'_\infty$ for the Ne atom. We observe that both $F_\infty^{\zPC}$ and  $F_\infty^{'\zPC}$ are quite close to PC and hPC for $r\le 2$ and $r\le 1.5$, respectively. At larger distances from the nucleus, the \zPC model 
behaves smoothly and approaches a positive constant \myr{ (in the asymptotic limit, the density decays exponentially, such that $s\rightarrow \infty$ and $z\rightarrow 1$, then $F_\infty^{\zPC}\approx a_1$ and $F_\infty^{'\zPC}\approx b_1$)}. 
\myr{
The asymptotic behaviors of $F_\infty$ and $F'_\infty$ in the SCE approach (see Eq. (\ref{eqI4})) using the conventional gauge \cite{gauge1}, is an open problem that is not investigated in this work. Here we only mention that $F_\infty^{MGGA}$ takes the maximum value allowed by locally satisfying the Lieb-Oxford bound; however, all the models shown in the figure (\zPC, hPC, PC, and MGGA) satisfy the Lieb-Oxford bound \cite{LObound}, which in the SCE limit becomes $W_\infty \ge -1.68 \int d\R n(\R)^{4/3}$.  
}

Then, we considered the total energies of atoms as computed using
the genISI2 ACII approach \cite{genisi2}.
In Tab. \ref{tab:Atoms} we report the results of the TE18 benchmark \cite{hfac24}, which represents the H-Ar atoms. We use the nearly complete 5ZaPa-NR-CV basis-set \cite{ranas15}; the reference data
is the CCSD(T) in the same basis-set\cite{hfac24}. (For more computational details, see Sec. \ref{sec6}.)
We considered effective exact-exchange orbitals \cite{lhf1,lhf2}.
The results show that while GL2 strongly overestimates the \myr{ total} energy, genISI2 gives an overall accuracy of only 3-4 mHa. Such accuracy is even superior to the one of HFAC24 functional (5 mHa)\cite{hfac24}.
Moreover, the \zPC results with MAE=3.4 mHa are slightly better than the hPC ones. 
Note that for the same benchmark, MP2 has MAE of 15.8 mHa \cite{hfac24}.
\myr{For comparison with other popular functionals, see Table 2 of Ref. \cite{hfac24}.}

\begin{table}
  \caption{\Dabel{tab:Atoms}Reference CCSD(T) total energy (in mHa) and the errors (approx.$-$reference) of GL2, genISI2-hPC,genISI2-\zPC and genISI2-TPSS functionals for the TE18 test. 
  The mean error (ME) and MAE (in mHa) are also reported.
  }
\begin{tabular}{lrrrrr}
\hline
\hline
   &      Ref.   &     GL2      &    genISI2       &   genISI2    & genISI2   \\
   &              &          &       hPC    &      \zPC   & TPSS \\
   \hline
H  & 	-500.000 & 	      {\bf 0.004} & 	   {\bf 0.004} & 	   {\bf 0.004} & 	  {\bf  0.004} \\ 
He  & 	-2903.233 & 	  -5.816 & 	   0.590 & 	  {\bf -0.073} & 	   1.227 \\ 
Li  & 	-7477.612 & 	  -4.064 & 	  {\bf -0.241} & 	  -0.656 & 	   0.366 \\ 
Be  & 	-14666.741 & 	 -30.200 & 	  -1.987   & 	  -3.835 & 	  {\bf -0.845} \\
B  & 	-24652.573 & 	 -28.964 & 	   {\bf 0.367} & 	  -2.858 & 	   1.646 \\ 
C  & 	-37842.983 & 	 -30.189 & 	   1.860 & 	  {\bf -1.570} & 	   3.687 \\ 
N  & 	-54586.561 & 	 -32.770 & 	   2.580 & 	  {\bf -0.772} & 	   5.011 \\ 
O  & 	-75062.803 & 	 -45.726 & 	   4.156 & 	  {\bf 0.487} & 	   7.048 \\ 
F  & 	-99727.414 & 	 -62.887 & 	   3.170 & 	  {\bf -0.217} & 	   7.076 \\ 
Ne  & 	-128929.226 & 	 -82.816 & 	   {\bf 0.083} & 	  -2.599 & 	   5.177 \\ 
Na  & 	-162246.996 & 	 -67.001 & 	   1.582 & 	   {\bf 0.371} & 	   6.591 \\ 
Mg  & 	-200045.220 & 	 -80.468 & 	  -3.013 & 	  -3.386 & 	   {\bf 2.445} \\ 
Al  & 	-242337.543 & 	 -78.090 & 	  {\bf -2.280} & 	  -2.421 & 	   3.158 \\ 
Si  & 	-289349.212 & 	 -82.588 & 	  -4.075 & 	  -3.873 & 	   {\bf 1.544} \\ 
P  & 	-341248.082 & 	 -91.377 & 	  -7.910 & 	  -7.274 & 	  {\bf -2.017} \\ 
S  & 	-398097.957 & 	-100.467 & 	  -7.724 & 	  -6.581 & 	  {\bf -1.509} \\ 
Cl  & 	-460135.451 & 	-115.610 & 	 -11.024 & 	  -9.307 & 	  {\bf -4.275} \\ 
Ar  & 	-527527.142 & 	-134.179 & 	 -16.634 & 	 -14.311 & 	  {\bf -9.263} \\ 
\hline
ME	&&	-59.623  & 	-2.250  & 	-3.271   &	{\bf 1.504}  \\
MAE	&&	59.623  & 	3.849  & 	{\bf 3.366}  & 	3.494  \\
    \hline
\end{tabular}
\end{table}

\subsection{Two-electron model density:}
\Dabel{sec32}
\begin{figure}[htb]
\includegraphics[width=\columnwidth]{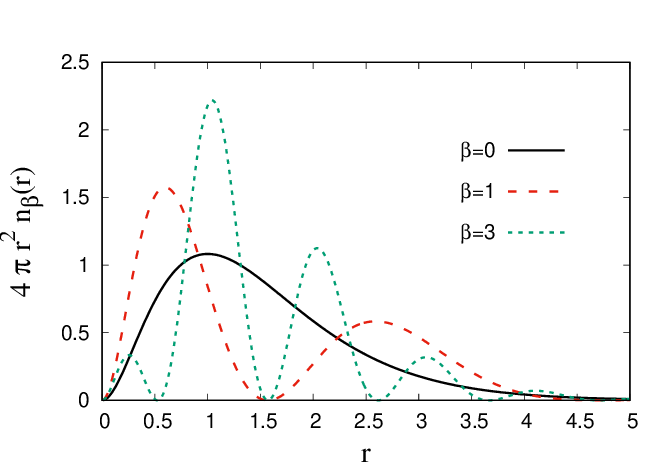}
\caption{The radial density $4\pi r^2 n_\beta(r)$ of the two-electron model density in  Eq. (\ref{eqne15}), versus the 
radial distance $r$, for several values of the $\beta$ parameter ($\beta=0$, 1 and 3).
}
\Dabel{f5}
\end{figure}
%
%
\begin{figure}[htb]
\includegraphics[width=\columnwidth]{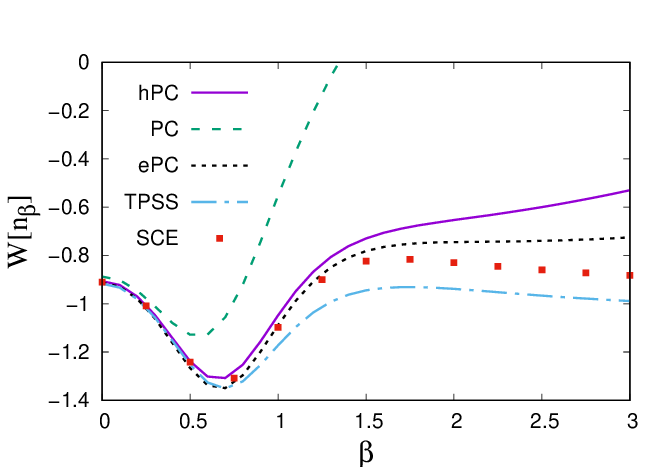}
\includegraphics[width=\columnwidth]{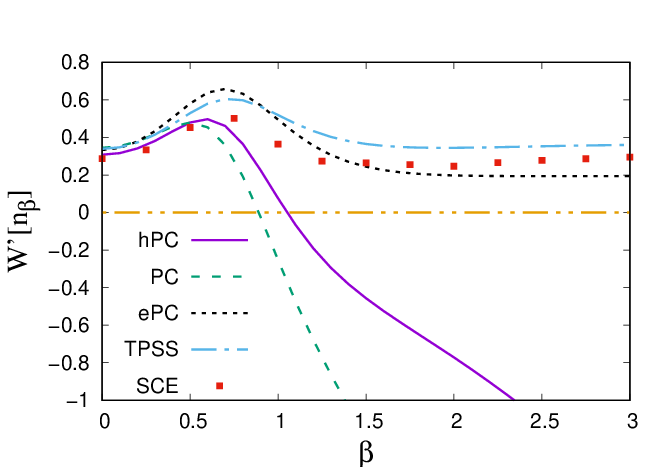}
\caption{The strong-interaction functionals $W[n_\beta]$ (upper panel) and $W'[n_\beta]$ (lower panel) versus the $\beta$ parameter for the two-electron model density of Eq. (\ref{eqne15}).}
\Dabel{f6}
\end{figure}
%
Let us consider the following $2 e^-$ spherical density 
\begin{equation}
n_\beta(r)=\frac{4(\beta^2+1)^3}{(\beta^6+3\beta^4+2)\pi}e^{-2r} \cos^2(\beta r),
\Dabel{eqne15}
\end{equation}
such that for any value of $\beta$, the normalization condition is  
$\int_0^\infty dr 4 \pi r^2 n_\beta(r) =2$. Note that when $\beta=0$ then 
the 2$e^-$ exponential density is found. In Fig. \ref{f5}, we plot this density for several values of the $\beta$ frequency.

Note that the SCE values were computed using the 
method presented in Ref. \onlinecite{seidl1999strong}. 

This model system is a very hard test for the PC and hPC models. 
As shown in Fig. \ref{f6}, PC and hPC start to fail even at relatively small values of $\beta$. 
In particular for $W'[n_\beta]$, PC and hPC give the wrong sign
for $\beta>1$.
On the other hand, both \zPC and TPSS models are 
remarkably accurate, \myr{ validating} our construction for $F_1(s)$ and $F'_1(s,\zeta)$,
\myr{ see Eqs. (\ref{eqne3}) and (\ref{eqne7}). 
}

\subsection{The $s$ and $p$ hydrogenic shells:}
\Dabel{sec35}
As another example, let us consider the hydrogenic orbitals $\psi_{nlm}(\R,Z)=R_{nl}(r,Z)Y_{lm}(\theta,\phi)$, 
that were used to investigate the nuclear \cite{hydrogenicnucleus} and asymptotic regions \cite{hydrogenicKS} in atoms, as well as the large atoms in the strong-interaction limit \cite{daas2023lo}.
Here $R_{nl}(r,Z)$ are the normalized radial functions, $Z$ is the nuclear charge, $Y_{lm}(\theta,\phi)$ are the spherical harmonics, and $n, l, m$ are the principal, angular momentum and azimuthal quantum numbers, respectively. The density of the fully filled $nl$ shell is
\begin{equation}
\rho_{nl}(r,Z)=\frac{(2l+1)}{2\pi}|R_{nl}(r,Z)|^2,
\Dabel{eqne17}
\end{equation}
and we consider neutral $s$ ($l=0,\;Z=N=2$) and $p$ ($l=1,\;Z=N=6$) shells. 

%
\begin{figure}[h]
\includegraphics[width=\columnwidth]{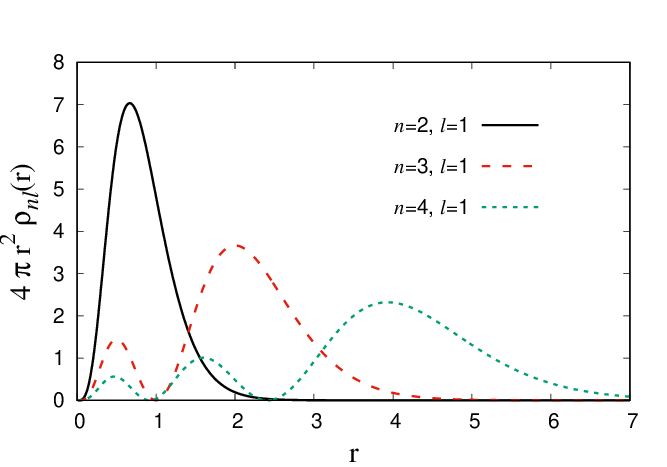}
\caption{The radial densities of the fully filled $p$-shells (see Eq. (\ref{eqne17})) versus the radial distance $r$. The area under each curve is the number of electrons ($N=6$).
}
\Dabel{f10}
\end{figure}
%
%
\begin{figure}[h]
\includegraphics[width=\columnwidth]{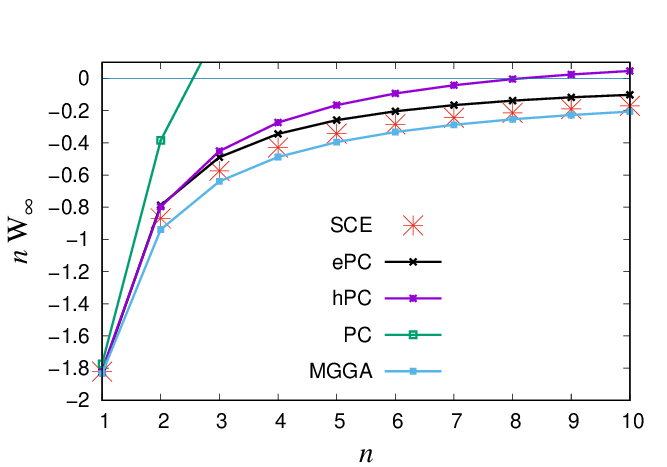}
\includegraphics[width=\columnwidth]{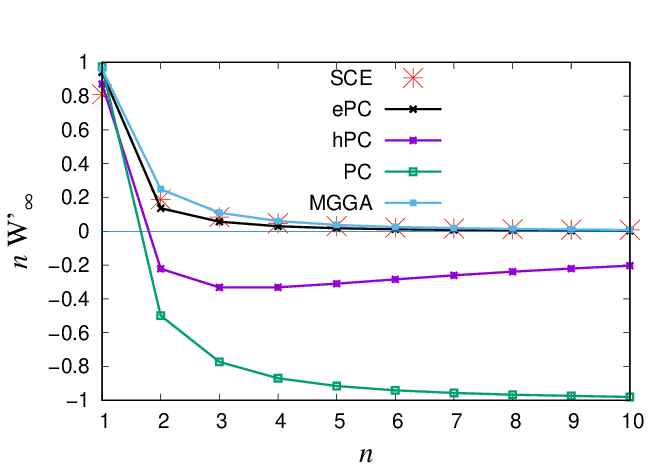}
\caption{$W_\infty$ (upper panel) and $W'_\infty$ (lower panel) multiplied with the principal quantum number $n$, versus $n$, for the \myr{ $s$-hydrogenic neutral shells (see Eq. (\ref{eqne17}) with $l=0, \;Z=N=2$). }
}
\Dabel{f11}
\end{figure}
%
%
\begin{figure}[h]
\includegraphics[width=\columnwidth]{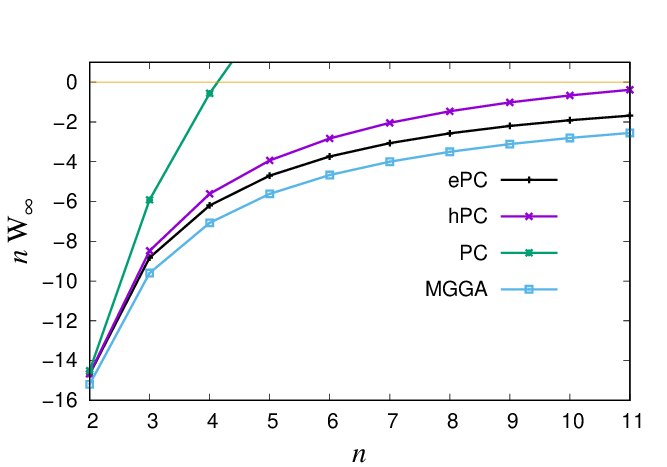}
\includegraphics[width=\columnwidth]{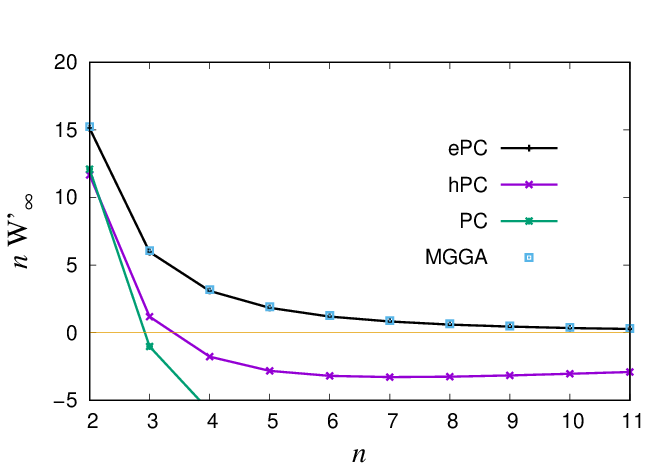}
\caption{Same as Fig. \ref{f11}, but for the \myr{ $p$-hydrogenic neutral shells  (see Eq. (\ref{eqne17}) with $l=1, \;Z=N=6$).} 
}
\Dabel{f12}
\end{figure}
%
In Fig. \ref{f10}, we show the radial densities ($4\pi r^2 \rho_{nl}$) for the $p$-shells ($l=1$) and $n=2, 3$ and 4. We observe that when $n$ increases, the density becomes more diffuse, similar to the real valence densities from pseudopotential calculations. 

In Figs. \ref{f11} and \ref{f12} we report the strong-interaction functionals $W_\infty$ and $W'_\infty$ as a function of the principal quantum number $n$, for the $s$- and $p$-shells, respectively. In case of the $s$-shell, we also compute the SCE values \cite{seidl1999strong}. 

The \zPC and meta-GGA models are accurate, yielding the correct behavior for all values of $n$. 
On the other hand, the hPC and PC functionals show an incorrect sign for increasing $n$, as also in the two-electron model density in the previous
subsection.
The hPC improves considerably over the PC, especially for $W_\infty$, which becomes positive only at very large $n$. 

The $s$- and $p$- hydrogenic shell models can be of interest to describe atoms with pseudopotential. In fact, in this case the density at the atomic core can be zero, and
the results presented here show that while PC and hPC can give even a wrong sign in those cases, this is not the case for \zPC.

\subsection{Quasi-two-dimensional infinite barrier model (quasi-2D IBM):}
\Dabel{sec33}
%
\begin{figure}[htb]
\includegraphics[width=\columnwidth]{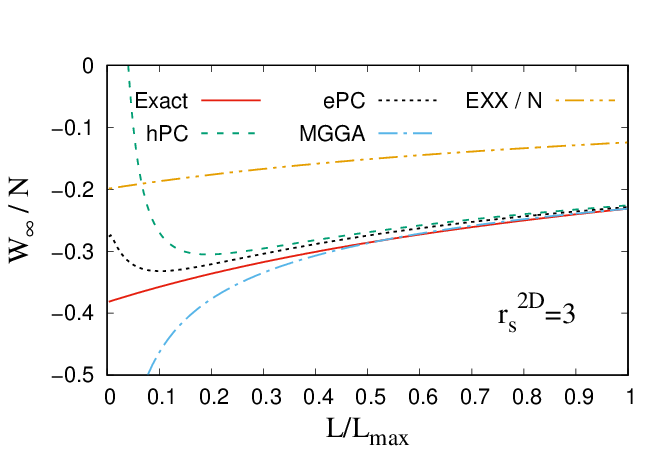}
\includegraphics[width=\columnwidth]{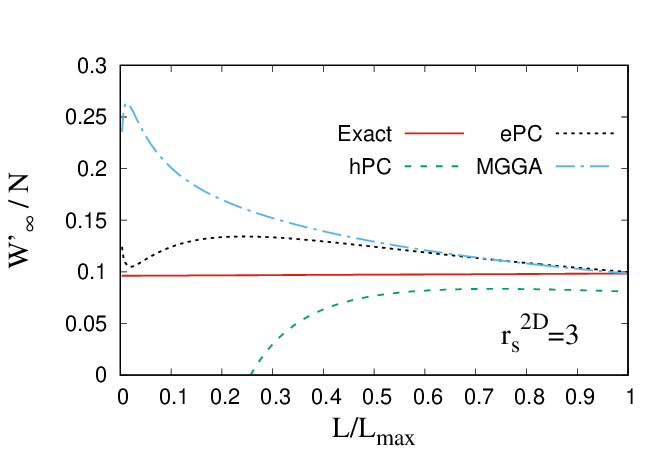}
\caption{$W_\infty(L)/N$ (upper panel) and $W'_\infty(L)/N$ (lower panel)
from hPC, \zPC and MGGA models, for the quasi-2D IBM with the 2D bulk parameter
$r_s^{2D}=3$. The reference curves are taken from Ref. \cite{genisi2}.  
}
\Dabel{f7}
\end{figure}
Let us consider the quasi-2D IBM quantum well
of thickness $L$ in the $z$-direction
\cite{pollack2000evaluating,karimi2014three,
constantin2016simple,kaplan2018collapse,horowitz2023construction}.
The true 2D uniform electron gas limit is recovered by shrinking
the $z$-coordinate, keeping fixed the total number of electrons
per unit area ($n^{2D}$).
The quasi-2D regime is obtained when
$L \leq \sqrt{3/2}\pi r_s^{2D}=L_{\rm{max}}$ \cite{pollack2000evaluating},
being equivalent to a non-uniform
scaling in one dimension (i.e. $n^z_{\lambda}(x,y,z)=\lambda n(x,y,\lambda z)$, with
$\lambda=L_{max}/L$).
 
Most of the semilocal functionals fail severely for the quasi-2D IBM, 
which represents a dimensional crossover, from 3D to 2D. Recently, we 
have shown that the ACII functionals are remarkably accurate for the quasi-2D IBM
when the ``exact'' expressions for $W_\infty(L)/N$ and $W'_\infty(L)/N$ are used,
which were constructed from a simple interpolation between the 2D and 3D expressions 
(see Eqs. (21) and (23) of Ref. \cite{genisi2}). 

In Fig. \ref{f7}, we show the performance of the \zPC model for the quasi-2D IBM with the 2D bulk parameter
$r_s^{2D}=3$. Note that similar results have been also found for
$r_s^{2D}=1$ and $r_s^{2D}=5$. We observe that $W^{\zPC}_\infty(L)/N$ is remarkably accurate 
for $L/L_{\rm{max}}\ge 0.1$, outperforming both hPC and MGGA. Note that 
$W_\infty(L)^{hPC}/N \le E_x^{EXX} /N$ at $L/L_{\rm{max}}\le 0.1$, failing badly and even becoming positive 
at $L/L_{\rm{max}}\le 0.05$. Also $W_\infty(L)^{MGGA}/N$ diverges when $L/L_{\rm{max}}\rightarrow  0$.

Moreover, \zPC also improves significantly over MGGA for $W'_\infty(L)/N$, while hPC 
fails drastically. 
Overall, we think that the \zPC performs encouragingly for the dimensional crossover, and it can 
be safely used in real quasi-2D calculations (e.g. two-dimensional materials 
\cite{8vvn-k9p3} and interfaces \cite{interfaces1}), where 
the strength of the quasi-2D regime is usually comparable with the case of quasi-2D IBM 
when $L/L_{\rm{max}}\rightarrow  0.3$.

\subsection{Perturbed uniform electron gas (UEG):}
\Dabel{sec34}
%
\begin{figure}[h]
\includegraphics[width=\columnwidth]{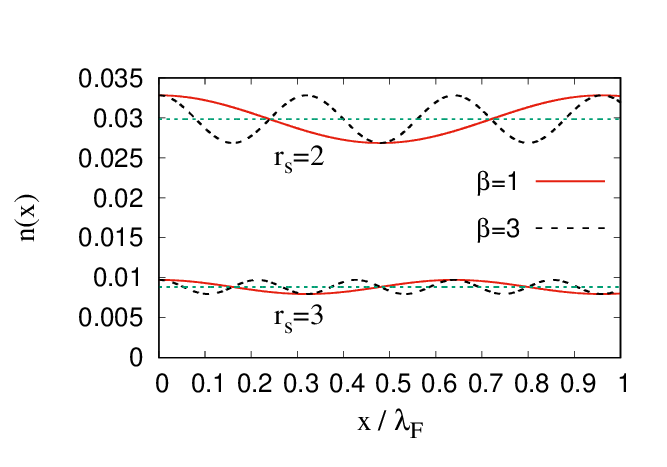}
\caption{The density $n(x)$ (see Eq. (\ref{eqne16})) versus the scaled distance $x /\lambda_F$ for bulk parameters $r_s=2$ and 3, and 
for several values of the frequency $\beta$.  
}
\Dabel{f8}
\end{figure}
%
Now let us consider a perturbed UEG in the $x$-direction, with the density
\begin{equation}
n(x)=n_0+\delta n(x)=n_0\Big[ 1+ A \cos(\beta x) \Big],
\Dabel{eqne16}
\end{equation}
where $n_0=3/(4\pi r_s^3)$ and we take $A=1/10$, then the perturbation 
density $\delta n(x,y,z)$ is small. Such a system is translationally invariant in the $(yz)$-plane.
In Fig. \ref{f8}, we show  $n(x)$ versus $x$ for bulk parameters $r_s=2$ and 3, and
for $\beta=1$ and $3$, respectively. Note that at $\beta=1$ the density varies slowly, while at $\beta=3$ 
there are small-amplitude, short-wavelength density waves, where the kinetic energy density 
is $\tau \rightarrow \tau^{TF}+\tau^W$ \cite{WJones_1971}, such that the meta-GGA ingredient is
$$z=(5s^2/3)/\big(1+(5s^2/3)\big).$$

%
\begin{figure}[h]
\includegraphics[width=\columnwidth]{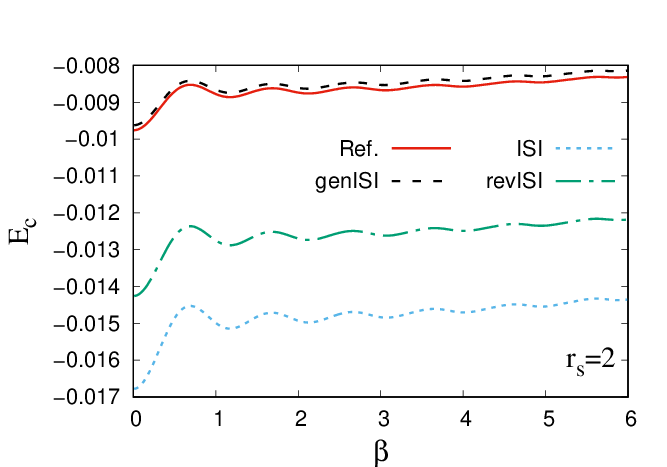}
\includegraphics[width=\columnwidth]{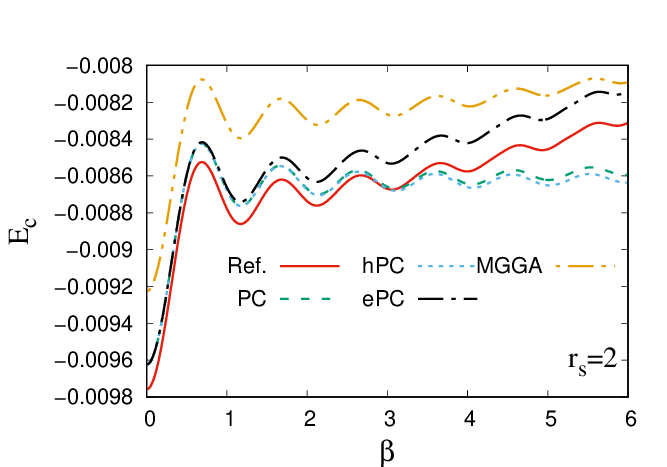}
\caption{Upper panel: ISI, revISI and genISI correlation energies per unit area $E_c$, for the perturbed UEG of Eq. (\ref{eqne16}), as a function 
of the frequency $\beta$. The reference is the TPSS correlation energy. The ACII functionals use the \zPC model. 
\newline
Lower panel: genISI correlation energy per unit area $E_c$ computed using PC, hPC, \zPC, and meta-GGA models for the strong-interaction functionals, for the perturbed UEG of Eq. (\ref{eqne16}), as a function
of the frequency $\beta$. The reference	is the TPSS correlation	energy. 
}
\Dabel{f9}
\end{figure}
%
We take $x$ in the interval $0\le x/\lambda_F \le 1$, where $\lambda_F=2\pi/k_F$ is the bulk Fermi wavelength, with $k_F=(3\pi^2 n_0)^{1/3}$ 
being the Fermi bulk wavevector, and $r_s=2$. For such a system, the TPSS correlation energy is very accurate,
and we consider it as the reference correlation. On the other hand, the GL2 correlation diverges so the ACII correlation functionals depend only on $E_x^{EXX}$, $W_\infty$ and $W'_\infty$.
 
First, we compute the ISI, revISI and genISI correlation energies per unit area $E_c$, within the \zPC model, and the 
results are reported in the upper panel of Fig. \ref{f9}. ISI and revISI significantly underestimate the correlation energy, whereas the genISI, which has been constructed from the UEG reference system, achieves good 
accuracy. We have also obtained similar patterns in the cases of UEG and jellium clusters \cite{constantin2023adiabatic}.

%
%
Finally, in the lower panel of Fig. \ref{f9} we show the genISI correlation energy per unit area $E_c$ computed using PC, hPC, \zPC, and meta-GGA models, 
such that we can make a fair comparison between these models. \zPC, PC and hPC are accurate, better than the meta-GGA that induces a slightly higher correlation energy. 

\subsection{$H_2$ dissociation}
\Dabel{sec36}
%
%
\begin{figure}[h]
\includegraphics[width=\columnwidth]{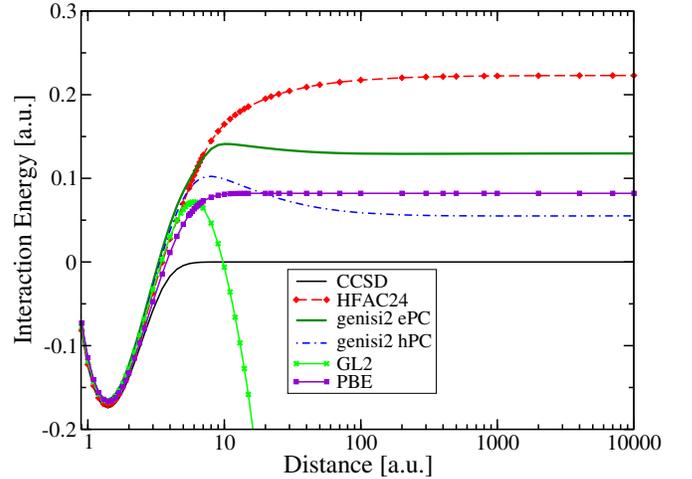}
\caption{ H$_2$ dissociation curve within a spin-restricted formalism.
GL2 and genISI2 using EXX orbitals, while the HFAC24 is based on HF orbitals. 
The exact and PBE curves are also shown.
}
\Dabel{fh2}
\end{figure}
%
In Fig. \ref{fh2}, we show the H$_2$ dissociation curve within a
spin-restricted formalism, a system that becomes a prototype example for the static correlation present in stretched molecular bonds and the dissociation limit 
\cite{weimer,zhang21,cohen08,Mori-Sanchez.Cohen.ea:Discontinuous.2009,Cohen.Mori-Sanchez.ea:Second-Order.2009,Janesko.Proynov.ea:Practical.2017,science21,genisi2}. The performance of the ACII XC functionals for this system has been recently investigated, e.g. see Refs. \cite{genisi2,hfac24,smiga2022selfconsistent}, such that here we comment only on the comparison between genISI2-hPC and genISI2-\zPC, computed with EXX orbitals. They agree closely until $R\approx 7$ bohr, where the use of hPC gives a repulsive bump, which is mostly attenuated in the case of \zPC curve. At the dissociation limit $D=E[H_2]-2E[H]$ is closer to exact for hPC than for \zPC. The better behavior of the hPC is due to an error cancellation between the positive and negative regions of $F_\infty^{'hPC}$, such that $0=W_\infty^{'SCE}\le W_\infty^{'hPC}\le W_\infty^{'\zPC}$. 
We also observe that genISI2-\zPC is superior in this limit to the HFAC24 functional \cite{hfac24}.

Obtaining $W_\infty^{'\zPC}=0$ for the spin-restricted hydrogen atom with half spin-up and half spin-down electrons, H(1/2,1/2), is not possible at the meta-GGA level,
with the enhancement factor non-negativity condition, as for this system $z=1$ and $\zeta=0$, like
all two-electrons atoms, where $W_\infty^{'\zPC}$ is not zero. Approaches beyond meta-GGAs are required.

\myr{Concerning $W_\infty$, we have that,  using exact orbitals (instead of EXX), $W_\infty$/2 = -0.3128, -0.3292, -0.3124, -0.3639 a.u.,    
for PC, hPC, ePC and TPSS, respectively, to be compared with the exact $W_\infty$/2=-0.3125 a.u.
Thus only PC and ePC give very accurate values.}

\section{Application to {\bf atomization energies and ionization potentials}:}

\subsection{Optimized hybrid orbitals:}
\Dabel{ophy}
The ACII XC methods are pure KS DFT approach, so that they must be  based on
the optimized effective potential (OEP) self-consistent KS orbitals.\cite{smiga2022selfconsistent},
The impact of the ground-state KS orbitals over the different ACII ingredients
has been investigated in Ref. \onlinecite{genisi2} and it was found that the GL2 terms strongly depend on the KS orbitals (in particular on the KS gap).
 
Self-consistent correlated OEP calculations, although possible \cite{smiga2022selfconsistent}, are numerically cumbersome, thus exchange-only OEP (OEPx) orbitals are usually used for the ACII functionals \cite{genisi2}, because they are close to the exact KS orbitals \myr{ for finite systems, such as atoms \cite{genisi2} and small molecules, where the correlation potential decays much faster in vacuum than the OEPx potential \cite{asymptotics1,PitarkePRL25}.} Nevertheless, OEPx orbitals are  quite challenging for large systems, too. 
In this work, we propose a simple optimized hybrid approach \myr{(in the generalized Kohn-Sham scheme)} in place of OEP, in order to compare hPC and \zPC in standard molecular benchmarks.

%
\begin{figure}[h]
\includegraphics[width=\columnwidth]{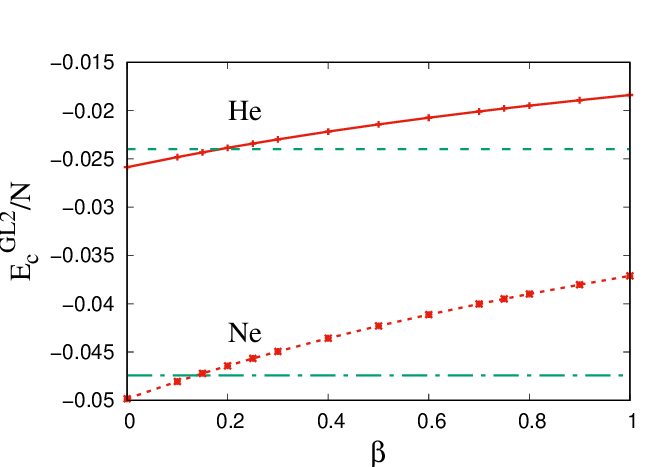}
\caption{The GL2 correlation per electron $E_c^{GL2}/N$ computed using the hybrid PBE orbitals with a fraction $\beta$ of non-local Hartree-Fock \myr{ -like} exchange, for He and Ne atoms. Also shown are the reference GL2 correlation energies (taken from Ref. \onlinecite{genisi2}).  
}
\Dabel{fophy1}
\end{figure}
%
We consider the self-consistent orbitals obtained from the hybrid PBE XC functional \cite{pbe0}  
%
%
\begin{equation}
E_{xc}[n](\beta)=\beta E_x^{HF}[n]+(1-\beta)E_x^{PBE}[n]+E_c^{PBE}[n],
\Dabel{eqT1}
\end{equation}
where the fraction $\beta$ of Hartree-Fock (HF) \myr{ -like} exchange $E_x^{HF}[n]$ needs to be optimized to obtain the exact GL2 correlation energy (see Table III of Ref. \onlinecite{genisi2}).
In this hybrid scheme, the corresponding GL2 correlation energy is:
%
%
\begin{eqnarray}
&& E_c^{GL2}[\{ \phi_a,\epsilon_a\}]=E_c^{MP2}[\{ \phi_a,\epsilon_a\}]-\nonumber\\
&& (1-\beta) \sum_{i,b}\frac{|\langle \phi_i |\hat{v}_x^{PBE}-\hat{v}_x^{HF}| \phi_b\rangle|^2}{\epsilon_b-\epsilon_i},
\Dabel{eqT2}
\end{eqnarray}
where $i,b$ indicate the occupied and virtual orbitals, respectively; and $\hat{v}_x^{PBE},\hat{v}_x^{HF}$ are the local PBE and non-local HF \myr{ -like} exchange potentials, respectively. 

In Fig. \ref{fophy1} we show $E_c^{GL2}$ as a function of the parameter $\beta$, for He and Ne atoms. The horizontal lines represent the exact GL2 correlation energies \cite{genisi2}. We observe that in both cases, for $\beta\approx 0.15$, the GL2 correlation is  close to exact. Thus, for molecular calculations, we use the orbitals from the functional in Eq. \eqref{eqT2}, with $\beta=0.15$ (for simplicity we name this functional PBE15).

\subsection{Atomization energies and ionization potentials }
\label{secaeip}

\begin{table}[ht]
    \caption{ \Dabel{tabae6} Reference atomization energies (in kcal/mol) and the errors (approx.-reference) of PBE, GL2, genISI2-hPC and genISI2-\zPC functionals for the AE6 test. The mean error (ME) and MAE (in kcal/mol) and MARE (in \%) are also reported. PBE15, GL2 and genISI2 all use PBE15    self-consistent orbitals. Reference values are from Ref. \cite{haunschild2012theoretical} and the basis-set is 5ZaPa-NR-CV\cite{ranas15}.}
    \begin{tabular}{lrrrrrr}
    \hline\hline
        &    Ref.     &    PBE     & PBE15    &      GL2                &  genISI2   &    genISI2  \\
        &             &            &           &                        &      hPC       &       \zPC           \\
\hline         
SiH$_4$ &	  324.59 & 	  -10.93 & 	   -9.71 & 	   13.43 & 	   -7.43 & 	   -6.81 \\ 
SiO  &	  193.40 & 	    3.33 & 	   -4.67 & 	   91.50 & 	   30.39 & 	   30.86 \\ 
S$_2$  & 	  103.90 & 	   11.52 & 	    6.59 & 	   46.68 & 	   17.50 & 	   17.30 \\ 
C$_2$H$_2$O$_2$ &	  635.33 & 	   27.91 & 	   10.46 & 	  159.42 & 	   13.58 & 	   22.22 \\ 
C$_3$H$_4$  &	  705.28 & 	   15.57 & 	    7.61 & 	   97.86 & 	   -0.63 & 	    7.70 \\ 
C$_4$H$_8$  & 		 1150.12 & 	   17.41 & 	   11.02 & 	  112.74 & 	  -12.96 & 	   -1.03 \\ 
\hline
ME &	&	   10.80  & 	    3.55  & 	   86.94  & 	    6.74  & 	   11.71  \\
MAE &	&	   14.45  & 	    8.34  & 	   86.94  & 	   13.75  & 	   14.32  \\
MARE &	&	    4.05  & 	    2.57  & 	   24.19  & 	    6.37  & 	    6.56  \\
 \hline\hline
    \end{tabular}
\end{table}
\begin{table}[ht]
    \caption{\Dabel{tab:ip} Reference ionization potential(Ref.), and the errors  obtained from PBE, PBE15, GL2, genISI2-hPC and genISI2-\zPC, for the G21IP benchmark test (but hydrogen atom) \cite{g21ip1,g21ip2,g21ip3}. The last rows show the error statistics; the mean error (ME), MAE and MARE (in \%). All the results are in kcal/mol. PBE15, GL2 and genISI2 all use PBE15
    self-consistent orbitals. Reference values are from Ref. \cite{g21ip3} and the basis-set is 5ZaPa-NR-CV\cite{ranas15}.}
        \begin{tabular}{lrrrrrr}
    \hline \hline
        Sys. & Ref. & PBE & PBE15 &  GL2   & genISI2  & genISI2  \\ 
             &       &       &    &         & hPC        & \zPC        \\ 
             \hline 
Li &	  123.30 & 	    5.58 & 	    5.33 & 	    1.17 & 	    1.08 & 	    1.34 \\ 
Be &	  214.90 & 	   -7.37 & 	   -7.73 & 	    9.17 & 	   -2.99 & 	   -2.14 \\ 
B &	  190.40 & 	    9.64 & 	    9.25 & 	    1.99 & 	    1.19 & 	    2.00 \\ 
C &	  259.60 & 	    6.54 & 	    6.15 & 	    3.47 & 	    0.24 & 	    0.52 \\ 
N &	  335.30 & 	    4.39 & 	    3.97 & 	    6.03 & 	    0.72 & 	    1.13 \\ 
O &	  313.80 & 	   10.48 & 	    7.47 & 	   12.86 & 	   -0.86 & 	   -0.07 \\ 
F &	  401.70 & 	    5.41 & 	    2.44 & 	   18.78 & 	    1.25 & 	    1.84 \\ 
Na &	  118.50 & 	    5.10 & 	    4.17 & 	    1.40 & 	   -0.11 & 	   -0.26 \\ 
Mg &	  176.30 & 	   -0.76 & 	   -1.75 & 	    9.63 & 	   -0.20 & 	   -0.33 \\ 
Al &	  138.00 & 	    2.20 & 	    2.45 & 	    2.23 & 	   -0.70 & 	   -0.48 \\ 
Si &	  188.00 & 	    1.04 & 	    1.32 & 	    4.37 & 	    0.31 & 	    0.34 \\ 
P &	  241.90 & 	   -0.06 & 	    0.49 & 	    6.62 & 	    1.92 & 	    1.92 \\ 
S &	  239.00 & 	    1.51 & 	    1.12 & 	    6.97 & 	   -1.21 & 	   -1.29 \\ 
Cl &	  299.10 & 	   -0.03 & 	   -0.28 & 	   10.07 & 	    0.80 & 	    0.69 \\ 
CH$_4$ &	  296.34 & 	   -9.10 & 	   -7.53 & 	   11.06 & 	    0.47 & 	    1.61 \\ 
NH$_3$ &	  235.69 & 	   -0.22 & 	   -2.12 & 	   22.75 & 	   -0.62 & 	    0.26 \\ 
OH &	  300.92 & 	    5.20 & 	    2.40 & 	   22.17 & 	    1.40 & 	    2.24 \\ 
OH$_2$ &	  292.65 & 	   -0.89 & 	   -2.99 & 	   26.60 & 	    0.99 & 	    1.77 \\ 
FH &	  371.31 & 	    0.15 & 	   -2.37 & 	   28.46 & 	    3.15 & 	    3.77 \\ 
SiH$_4$ &	  255.39 & 	   -7.20 & 	   -5.67 & 	    6.07 & 	   -1.10 & 	   -0.95 \\ 
PH &	  234.11 & 	    1.93 & 	    2.54 & 	    4.99 & 	    1.08 & 	    1.10 \\ 
PH$_2$ &	  226.37 & 	    3.39 & 	    3.95 & 	    3.72 & 	    0.13 & 	    0.15 \\ 
PH$_3$ &	  227.82 & 	   -2.43 & 	   -3.18 & 	   13.77 & 	    0.86 & 	    0.75 \\ 
SH &	  239.30 & 	    0.04 & 	   -0.31 & 	   11.96 & 	    1.64 & 	    1.52 \\ 
HCl &	  294.46 & 	   -2.09 & 	   -2.19 & 	   14.43 & 	    3.09 & 	    2.95 \\ 
C$_2$H$_2$ &	  264.58 & 	   -4.72 & 	   -5.85 & 	   27.06 & 	   -0.28 & 	    1.30 \\ 
C$_2$H$_4$ &	  243.71 & 	   -5.03 & 	   -5.97 & 	   21.28 & 	   -2.02 & 	   -0.62 \\ 
N$_2$ &	  359.37 & 	   -4.98 & 	    0.27 & 	  -55.67 & 	  -13.19 & 	  -14.67 \\ 
O$_2$ &	  277.73 & 	    4.31 & 	    6.22 & 	  -22.40 & 	   -5.86 & 	   -6.26 \\ 
P$_2$ &	  242.85 & 	    2.00 & 	    3.75 & 	    0.85 & 	   -0.91 & 	   -0.91 \\ 
S$_2$ &	  215.74 & 	    1.26 & 	    3.37 & 	  -14.79 & 	   -9.42 & 	   -9.28 \\ 
Cl$_2$ &	  265.08 & 	   -8.87 & 	   -6.25 & 	   -2.95 & 	   -4.61 & 	   -4.60 \\ 
ClF &	  291.70 & 	   -7.41 & 	   -5.42 & 	   -8.38 & 	   -8.99 & 	   -8.96 \\ 
\hline
ME &	&	    0.27  & 	    0.21  & 	    6.23  & 	   -0.99  & 	   -0.72  \\
MAE &	&	    3.98  & 	    3.83  & 	   12.55  & 	    2.22  & 	    2.36  \\
MARE &	&	    1.68  & 	    1.65  & 	    4.58  & 	    0.84  & 	    0.88  \\
\hline\hline
    \end{tabular}
\end{table}

We employ the hPC and \zPC strong-interaction models, in the genISI2 ACII XC functional \cite{genisi2}, for two standard molecular benchmarks. Calculations have been done in a nearly complete basis-set (see Computational Details section) in order to avoid
basis-set inaccuracies for the calculation of the GL2 correlation energy \cite{hfac24}.

In Table \ref{tabae6} we report the atomization energies of molecules from the popular benchmark test AE6 \cite{lynch2003small,haunschild2012theoretical}. GL2 strongly overestimates the atomization energies of all AE6 molecules, with MAE$\approx 90$ kcal/mol, almost one order of magnitude worse than MP2 \cite{lynch2003small}. We recall that the main difference between GL2 and MP2 is given by the type of orbitals, 
\myr{ as the HOMO-LUMO gap from the PBE15 orbitals is considerably
smaller than the gap from the HF orbitals.}
On the other hand, genISI2 strongly improves over GL2, with MAE$\approx 14$ kcal/mol, which is close to the PBE one. Comparing genISI2-hPC and genISI2-\zPC, we found that on average they agree closely, both having MAE$\approx$14 kcal/mol and MARE$\approx$6 \%, despite the \myr{fact that the results} for individual molecules are different (for the molecules C$_2$H$_2$O$_2$, C$_3$H$_4$, and C$_4$H$_8$, they about 8-11 kcal/mol).
Note that results in Table~\ref{tabae6} differs from Tab. VI of Ref. \cite{genisi2} due to the different basis-set and different
reference energy. \myr{. For comparison with other popular functionals, see Table 2 of Ref. \cite{hfac24}.}

In Table \ref{tab:ip} we report the results for the ionization
potential benchmark G21IP test \cite{g21ip3}, but CO and CS, which are spin-contaminated\cite{hfac24}. Tab. \ref{tab:ip} shows that genISI2-hPC and genISI2-\zPC give similar results for all systems, with differences less than 0.8 kcal/mol . Consequently, they have similar error statistics, with MAE$\approx$2.3 kcal/mol and MARE$\approx$0.8 \%. These results are noticeably better than PBE15 (with MARE$\approx1.7$ \%) and GL2 (MARE$\approx4.6$ \%), as well as many other popular XC functionals \cite{g21ip1,g21ip2,hapbe}.   
Thus, the \zPC strong-correlation functionals retain
the accuracy of the hPC functional. \myr{For comparison with other popular functionals, see Table 2 of Ref. \cite{hfac24}.}

\myr{The} largest error is found for the $N_2$ molecule. In the cation state we found a small energy-gap which gives a very large GL2 energy, so that the energy difference with respect to the neutral ground-state is reduced: genISI2 largely corrects the GL2 underestimation, but not completely.

\section{Conclusions} 
\Dabel{sec5}
We have constructed the enhanced point-and-charge (\zPC) meta-GGA model for the strong-interaction functionals: $W_\infty[n]$, which represents the SCE limit of the adiabatic connection integrand $W_\infty[n]=\lim_{\alpha\rightarrow\infty}W_{xc,\alpha}[n]$ and $W'_\infty[n]$, which accounts for the electronic zero-point oscillations in the SCE limit. The \zPC model \myr{ resolves} several shortcomings of the existing approximations, and fulfills important exact conditions: it is exact for the hydrogen atom, recovers the PC-GE2, satisfies important inequalities, in particular the non-negativity of $W'_\infty[n]$, and it is spin-dependent only for systems with few electrons and rapidly-varying densities. In particular, \zPC is the first model that incorporates the full restoration of the PC-GE2, that should be relevant for Wigner crystals at low densities \cite{seidl2000density,seidl2000densitye}. 

In the case of systems with compact densities, as the atoms and molecules of Tables \ref{tab1},\ref{tab:Atoms},\ref{tabae6} and \ref{tab:ip}  
the \zPC model performs \myr{ somewhat}  better than 
the hPC state-of-the-art semilocal model.

For more diffuse densities, as the two-electron model density of Sec. \ref{sec32} and the $s$- and $p$-shells of hydrogenic orbitals presented in Sec. \ref{sec35}, the \zPC and TPSS are remarkably accurate, while hPC and PC fail badly, even changing their signs. 
The density of these model systems resembles the one of systems described with pseudopotentials (i.e. with a vanishing density at the atomic core). Thus the \zPC model can be expected to be 
very important in this case, and investigations in this direction are on the way.

The \zPC model also has a remarkable performance for the quasi-2D IBM model system (see Sec. \ref{sec33}), that represents a dimensional crossover from 3D UEG to 2D UEG where the density is still compact but with higher reduced gradients than in the case of atoms and molecules. As shown in Fig. \ref{f7}, all the other strong-interaction models are failing badly. These facts show that the \zPC model achieves a broader applicability compared with the existing strong-interaction semilocal models.    

The \zPC model can be further useful in the DFT development of XC functionals. Thus, its energy densities $w_\infty^{\zPC}(\R)$ and $w_\infty^{'\zPC}(\R)$, defined from
$W_\infty^{\zPC}[n]=\int d\R\; w_\infty^{\zPC}(\R)$ and $W_\infty^{'\zPC}[n]=\int d\R\; w_\infty^{'\zPC}(\R)$, may be of interest for local interpolations along the adiabatic connection, a method for constructing XC functionals that has gained recent investigations and good progress \cite{mirtschink2012energy,vuckovic2017interpolated,zhou2015construction,bahmannshellmodel,levyaugmentedenergydensity,giarrussoeneden1,giarrussoeneden2,vuckovic2016exchange,kooi2018local,constantin2019correlation,jana2023semilocal}, being represented on the entire Jacob's ladder \cite{Jacob_ladder}, from the semilocal GGA and meta-GGA levels \cite{constantin2019correlation,jana2023semilocal} to the most sophisticated XC functionals, depending on the occupied and unoccupied KS orbitals and eigenvalues as well as the comotion functions $f_i(\R)$ and the local normal modes $\omega_\mu(\R)$ \cite{kooi2018local}.  \\

\section*{Computational Details}
\Dabel{sec6}
For the atomic and molecular calculations of Tables \ref{tab:Atoms}, \ref{tabae6} and \ref{tab:ip}, and Figs. \ref{fh2} and \ref{fophy1},
we have used a modified version of the program package TURBOMOLE\cite{turbo2023} together with the  acmxc\cite{acmxc} script. The calculations have been performed within the resolution of identity (RI) approximation \cite{RI1,RI2} and the nearly complete 5ZaPa-NR-CV all-electron basis sets\cite{ranas15}, with the gridsize 4.
We used self-consistent hybrid PBE15 orbitals, as discussed above, or the Localized Hartree-Fock orbitals \cite{lhf1,lhf2} for the atoms in Tab. \ref{tab:Atoms}.

The SCE calculations of the atoms in Table \ref{tab1} have been performed with the Engel code \cite{engel1993accurate,Engel2003}, using accurate numerical OEP-EXX orbitals and densities, in a logarithmic grid. The calculations for Hooke's atoms, two-electron exponential densities, quasi-2D IBM, perturbed UEG and the $s$- and $p$-hydrogenic shells use exact densities. 

\section*{Acknowledgments}
F.D.S, E.F. and F.S. acknowledge the financial support from
PRIN project no. 2022LM2K5X, Adiabatic Connection for Correlation in Crystals ($AC^3$).
L.A.C. and F.D.S. acknowledge the financial support from
ICSC - Centro Nazionale di Ricerca in High Performance
Computing, Big Data and Quantum Computing, funded by European Union - NextGenerationEU - PNRR.

\section*{Data Availability}
The data that support the findings of this article are not publicly available upon publication because it is not technically feasible and/or the cost of preparing, depositing, and hosting the data would be prohibitive within the terms of this research project. The data are available from the authors upon reasonable request.

\twocolumngrid
\bibliography{gisi2}

\end{document}